\documentclass[preprint,showpacs,showkeys,preprintnumbers,amsmath,amssymb,aps]{revtex4-1}

\usepackage{bm,amssymb,amsmath,mathrsfs}
\usepackage[usenames]{color}
\usepackage[colorlinks]{hyperref}

\usepackage{dcolumn}

\def\half{{\textstyle\frac{1}{2}}}

\newcommand{\bc}{\begin{center}}
\newcommand{\ec}{\end{center}}
\newcommand{\bit}{\begin{itemize}}
\newcommand{\eit}{\end{itemize}}
\newcommand{\bq}{\begin{equation}}
\newcommand{\eq}{\end{equation}}

\begin{document}

\title{Generalizations of the McMillan map to $N$-body systems}

\author{S.~R.~Mane}
\email{srmane001@gmail.com}

\affiliation{Convergent Computing Inc., P.~O.~Box 561, Shoreham, NY 11786, USA}

\begin{abstract}
The McMillan map is a well-known example of a rational integrable system for one particle in a two-dimensional phase space.
An elegant recent paper presented a generalization of the McMillan map to an $N$-body system, 
for particles moving in $d$ space dimensions.
This paper presents some alternative generalizations (also completely integrable) of the McMillan map to $N$-body systems.
In all cases, the phase space is foliated by a biquadratic curve in the dynamical variables
(and a set of suitably chosen angular momentum variables). 
It is also demonstrated that the constraints to generalize the McMillan map to $N$-body systems are not trivial.
\end{abstract}



\maketitle

\setcounter{equation}{0}

In many cases, the analysis of dynamical systems lends itself naturally 
to the use of maps or so-called Poincar{\'e} sections.
An excellent exposition can be found, e.g., in the text by Dragt \cite{DragtBook}.
(The particle motion around the circumference of a synchrotron provides a natural setting for the use of maps.)
Of particular interest are maps which describe completely integrable systems,
both as approximations to real systems and for pedagogical interest in dynamical systems theory.
The McMillan map \cite{McMillan_map} is a fully integrable nonlinear system in a one-body phase space,
and the prototypical example of a so-called rational integrable system, 
i.e.~the equations are expressible as rational functions (ratio of two polynomials) of the dynamical variables.
See \cite{Iatrou_Roberts} and references therein for more details of rational integrable systems.
An elegant recent paper \cite{DanilovNagaitsev_PRSTAB2014}
has presented a generalization of the McMillan map to an $N$-body system, 
for particles moving in $d$ space dimensions.

This paper shows that the generalization in \cite{DanilovNagaitsev_PRSTAB2014}
is not unique, and other (also fully integrable) generalizations of the McMillan map to $N$-body systems are possible.
The present paper investigates some alternative models and the conditions for integrability, 
when generalizing the McMillan map to $N$-body systems.
The conditions are shown to be not trivial.
All of the models below are examples of measure-preserving (in fact symplectic)
rational integrable systems and the phase spaces are foliated 
by a suitably chosen set of angular momenta and biquadratic curves in the dynamical variables, as will be explained below.

Note that the analysis in \cite{DanilovNagaitsev_PRSTAB2014}
was with respect to motion in circular particle accelerators.
However, the McMillan map is completely general, and the material presented below deals exclusively with maps {\em per se},
without any specific reference to particle accelerators.

First some details of the McMillan map \cite{McMillan_map} are summarized.
The notation and terminology below mostly follows \cite{Iatrou_Roberts}.
For simplicity, it is assumed that all the dynamical variables have been scaled to be dimensionless.
The ``McMillan map'' is actually a family of nonlinear integrable maps in the $(x,p)$ plane 
\bq
\label{eq:mcmillam_map}
x^\prime = p \,,\qquad
p^\prime = -x - \frac{\beta p^2 +\epsilon p + \xi}{\alpha p^2 + \beta p + \gamma} \,.
\eq
Here $\alpha$, $\beta$, $\gamma$, $\epsilon$ and $\xi$ are constants and a prime denotes application of the map.
The above is an area-preserving rational family of mappings preserving the biquadratic foliation
\bq
\label{eq:foliation}
\alpha x^2 p^2 + \beta (x^2 p + x p^2) + \gamma (x^2 + p^2) + \epsilon xp + \xi (x + p) + K = 0\,.
\eq
Here $K$ is a parameter which parametrizes each invariant curve in the plane.
Note in passing that the general formalism allows $\alpha$, $\beta$, etc.~above to depend on the invariant $K$.
To avoid clumsy notation, this fact will be implicitly assumed below throughout.
Following \cite{DanilovNagaitsev_PRSTAB2014}, the individual particles in an $N$-body system
will be indexed by a subscript $i$, hence ``$x_i$'' and ``$p_i$.''
An overbar is employed to denote an average, i.e.~$\bar{f} = (1/N)\,\sum_{i=1}^N f_i$ for any quantity $f$.
In terms of the above notation, the map in \cite{DanilovNagaitsev_PRSTAB2014} is given by
\bq
x_i^\prime = p_i \,,\qquad
p_i^\prime = -x_i - \frac{\epsilon \,p_i}{\alpha\,\overline{p^2} + \gamma} \,.
\eq
The authors in \cite{DanilovNagaitsev_PRSTAB2014} have set $\beta=\xi=0$ and $\gamma=1$.
If $\gamma\ne 0$, we can scale 
$(\alpha,\beta,\gamma,\epsilon,\xi) \to (\alpha/\gamma,\beta/\gamma,1,\epsilon/\gamma,\xi/\gamma)$.
If $\gamma=0$, the proof of integrability in \cite{DanilovNagaitsev_PRSTAB2014} still works, but the motion may become unstable.
It is noted in \cite{DanilovNagaitsev_PRSTAB2014} that the angular momenta
$M_{ij} = x_ip_j - x_jp_i$ (where $i\ne j$) are all dynamical invariants. 
A set of $N-1$ invariants is given as follows, for $j = 1,\dots,N-1$
(see eq.~(7) in \cite{DanilovNagaitsev_PRSTAB2014})
\bq
\label{eq:angmtm_inv}
I_j = \sum_{i=1}^j (M_{i,j+1})^2 \,.
\eq
A further invariant is given by (see eq.~(13) in \cite{DanilovNagaitsev_PRSTAB2014})
\bq
\label{eq:dn_inv}
I = \alpha\,(\overline{xp})^2 + \epsilon\,\overline{xp} + \gamma\,(\overline{x^2} + \overline{p^2}) \,.
\eq
It was proved in \cite{DanilovNagaitsev_PRSTAB2014} that the set of $N$ invariants in
eqs.~\eqref{eq:angmtm_inv} and \eqref{eq:dn_inv} are in involution,
i.e.~their mutual Poisson Brackets vanish, hence the system is completely integrable.
The model is generalized in \cite{DanilovNagaitsev_PRSTAB2014} to treat the motion of $N$ particles in a $d$-dimensional space.
The invariant can be expressed in vector notation as
$I = \alpha\,(\overline{\bm{r}\cdot\bm{p}})^2
+\epsilon\,\overline{\bm{r}\cdot\bm{p}}
+\gamma\,(\overline{\bm{r}^2}+\overline{\bm{p}^2})$.
I note in passing a more succinct derivation of the invariance of the invariant $I$ 
presented in \cite{DanilovNagaitsev_PRSTAB2014}:
\bq
\begin{split} 
I^\prime &= \alpha\,(\overline{\bm{r}^\prime\cdot\bm{p}^\prime})^2
+\epsilon\,\overline{\bm{r}^\prime\cdot\bm{p}^\prime}
+\gamma\,(\overline{\bm{r}^{\prime 2}}+\overline{\bm{p}^{\prime 2}}) 
\\
&= \alpha\,\overline{\bm{p}\cdot\biggl(\bm{r} + \frac{\epsilon \bm{p}}{\alpha\,\overline{\bm{p}^2} + \gamma} \biggr)}^{\,2}
-\epsilon\,\overline{\bm{p}\cdot\biggl(\bm{r} + \frac{\epsilon \bm{p}}{\alpha\,\overline{\bm{p}^2} + \gamma} \biggr)}
+\gamma\, \overline{\bm{p}^2}
+\gamma\, \overline{\biggl(\bm{r} + \frac{\epsilon \bm{p}}{\alpha\,\overline{\bm{p}^2} + \gamma} \biggr)^2}
\\
&= \alpha\,
\biggl(\overline{\bm{r}\cdot\bm{p}}^2 
+ \frac{2\epsilon\,\overline{\bm{p}^2}\, \overline{\bm{r}\cdot\bm{p}}}{\alpha\,\overline{\bm{p}^2} + \gamma} 
+ \frac{\epsilon^2\,\overline{\bm{p}^2}^2}{(\alpha\,\overline{\bm{p}^2} + \gamma)^2} 
\biggr)
\\
&\quad
-\epsilon\,\biggl(\overline{\bm{r}\cdot\bm{p}} + \frac{\epsilon\,\overline{\bm{p}^2}}{\alpha\,\overline{\bm{p}^2} + \gamma} \biggr)
+\gamma\, \overline{\bm{p}^2}
+\gamma\, \biggl( \overline{\bm{r}^2} + \frac{2\epsilon\,\overline{\bm{r}\cdot\bm{p}}}{\alpha\,\overline{\bm{p}^2} + \gamma} 
+ \frac{\epsilon^2\, \overline{\bm{p}^2}}{(\alpha\,\overline{\bm{p}^2} + \gamma)^2} \biggr)
\\
&= I
+\biggl(2\epsilon\,\overline{\bm{r}\cdot\bm{p}} +\frac{\epsilon^2\,\overline{\bm{p}^2}}{\alpha\,\overline{\bm{p}^2} + \gamma}\biggr)
\underbrace{\biggl( \frac{\alpha\,\overline{\bm{p}^2} +\gamma}{\alpha\,\overline{\bm{p}^2} + \gamma} -1\biggr)}_{=0} \,.
\end{split}
\eq
Hence $I^\prime = I$ and the invariance is established.

Consider now, {\em provisionally}, a different interaction, also nonlinear,
where we square the sum of the momenta (as opposed to the sum of squares)
\bq
\label{eq:toomany}
x_i^\prime = p_i \,,\qquad
p_i^\prime = -x_i - \frac{\beta\, p_i^2 +\epsilon \,p_i +\xi}{\alpha\,\bar{p}^2 +\beta\,\bar{p} + \gamma} \,.
\eq
This contains too many terms. 
The angular momenta are {\em not} conserved, in general, under the application of the above map
\bq
\begin{split} 
M_{ij}^\prime &= x_i^\prime p_j^\prime - x_j^\prime p_i^\prime
\\
&= -p_i\biggl( x_j + \frac{\beta p_j^2 +\epsilon p_j +\xi}{\alpha\,\bar{p}^2+\beta\,\bar{p}+\gamma}\biggr)
+p_j\biggl( x_i + \frac{\beta p_i^2 +\epsilon p_i +\xi}{\alpha\,\bar{p}^2+\beta\,\bar{p}+\gamma}\biggr)
\\
&= M_{ij} + \frac{\beta p_ip_j  -\xi}{\alpha\,\bar{p}^2+\beta\,\bar{p}+\gamma}\,(p_i-p_j) \,.
\end{split}
\eq
The angular momentum is conserved only if $\beta=0$ and $\xi=0$.
Note in passing that a similar derivation also explains why the map in 
\cite{DanilovNagaitsev_PRSTAB2014} also sets $\beta=\xi=0$.
The McMillan map permits nonzero values of $\beta$ and $\xi$ because it treats only $N=1$, so $i=j=1$ and $p_i-p_j=0$.
Hence we postulate the following model, 
where we generalize immediately to a model of $N$ particles moving in $d$ space dimensions.
With an obvious notation,
\bq
\label{eq:newmodel}
\bm{r}_i^\prime = \bm{p}_i \,,\qquad
\bm{p}_i^\prime = -\bm{r}_i - \frac{\epsilon\,\bm{p}_i}{\alpha\,\bar{\bm{p}}^2 + \gamma} \,.
\eq
Here the indices $i$ and $j$ run from 1 to $N$, and each vector $\bm{r}_i$ and $\bm{p}_i$ has $d$ components.
The new definition of the angular momenta is
$M_{ijk\ell} = (\bm{r}_i)_k(\bm{p}_j)_\ell - (\bm{r}_j)_\ell(\bm{p}_i)_k$.
Here $k$ and $\ell$ run from $1$ through $d$.
If $i=j$ then we must have $k\ne\ell$ and if $k=\ell$ then we must have $i\ne j$.
Note that the above definition includes terms such as 
$x_1p_{x2} - x_2p_{x1}$, which we do not not normally think of as angular momentum.
(The same remark applies to the definition of the angular momenta in
\cite{DanilovNagaitsev_PRSTAB2014}.)
It is straightforward to derive a set of $dN-1$ invariants analogous to those in
eq.~\eqref{eq:angmtm_inv}, say $\tilde{I}_\nu$ 
(where $\nu=1,\dots,dN-1$ and we require a sum with a messy collection of indices),
which will not be written out explicitly here.
To obtain the final invariant, we sum over $i$ 
to obtain map equations for the centroid $(\bar{\bm{r}}, \bar{\bm{p}})$
\bq
\label{eq:centroid}
\bar{\bm{r}}^\prime = \bar{\bm{p}} \,,\qquad
\bar{\bm{p}}^\prime = -\bar{\bm{r}} - \frac{\epsilon\,\bar{\bm{p}}}{\alpha\,\bar{\bm{p}}^2 + \gamma} \,.
\eq
This is effectively a single particle mapping analogous to eq.~\eqref{eq:mcmillam_map}.
It follows from eq.~\eqref{eq:foliation} that the invariant is
\bq
J = \alpha\,\bar{\bm{r}}^2 \bar{\bm{p}}^2 
+\epsilon\,\bar{\bm{r}}\cdot\bar{\bm{p}} 
+\gamma(\bar{\bm{r}}^2+\bar{\bm{p}}^2) \,.
\eq
Next, the Poisson Bracket of $J$ with any angular momentum vanishes,
because $J$ is symmetric under an interchange $(i,j,k,\ell) \leftrightarrow (j,i,\ell,k)$,
whereas $M_{ijk\ell}$ is antisymmetric, hence $\{ J, M_{ijk\ell} \} = 0$.
This establishes that $\{ J, \tilde{I}_\nu \}=0$ for all $\nu=1,\dots,dN-1$.
Hence the system is completely integrable.
It is immediate that we can generalize $\alpha$ to a vector:
\bq
\bm{r}_i^\prime = \bm{p}_i \,,\qquad
\bm{p}_i^\prime = -\bm{r}_i - \frac{\epsilon\,\bm{p}_i}{\overline{\bm{\alpha}\cdot\bm{p}}^2 + \gamma} \,.
\eq
The angular momenta $M_{ijk\ell}$ are all conserved, hence also the $\tilde{I}_\nu$.
The additional invariant is
\bq
\hat{J} = \overline{\bm{\alpha}\cdot\bm{r}}^2 \;\overline{\bm{\alpha}\cdot\bm{p}}^2 
+\epsilon\,\overline{\bm{\alpha}\cdot\bm{r}}\;\overline{\bm{\alpha}\cdot\bm{p}} 
+\gamma\,(\overline{\bm{\alpha}\cdot\bm{r}}^2 + \overline{\bm{\alpha}\cdot\bm{p}}^2) \,.
\eq
It is straightforward to derive the proof of invariance.
We cannot, however, generalize $\epsilon$ to a tensor $\bm{p}_i \to \sum_m \epsilon_{im} \bm{p}_m$,
because the angular momenta will not be map invariants.
For the map in \cite{DanilovNagaitsev_PRSTAB2014}, 
we can go one step further and generalize $\alpha$ to a symmetric tensor.
With an obvious notation, the map equations are
\bq
\bm{r}_i^\prime = \bm{p}_i \,,\qquad
\bm{p}_i^\prime = -\bm{r}_i - \frac{\epsilon\,\bm{p}_i}{\overline{\bm{p}\cdot\bm{\alpha}\cdot\bm{p}} + \gamma} \,.
\eq
The angular momenta $M_{ijk\ell}$ are map invariants, hence also the $\tilde{I}_\nu$.
The additional invariant is, written in an explicitly symmetric form
\bq
\hat{I} = \overline{\bm{r}\cdot\bm{\alpha}\cdot\bm{r}} \;\overline{\bm{p}\cdot\bm{\alpha}\cdot\bm{p}}
+\half\epsilon\,(\overline{\bm{r}\cdot\bm{\alpha}\cdot\bm{p}} +\overline{\bm{p}\cdot\bm{\alpha}\cdot\bm{r}})
+\gamma\,( \overline{\bm{r}\cdot\bm{\alpha}\cdot\bm{r}} + \overline{\bm{p}\cdot\bm{\alpha}\cdot\bm{p}} ) \,.
\eq
The above are all models of completely integrable systems.
Let us consider some other related models and investigate the invariants.
Consider a map with constant vectors $\bm{\xi}_i$ as follows:
\bq
\bm{r}_i^\prime = \bm{p}_i \,,\qquad
\bm{p}_i^\prime = -\bm{r}_i - \frac{\epsilon\,\bm{p}_i+\bm{\xi}_i}{\alpha\,\bar{\bm{p}}^2 + \gamma} \,.
\eq
Then
\bq
\begin{split}
M_{ijk\ell}^\prime &= (\bm{r}_i)_k^\prime(\bm{p}_j)_\ell^\prime - (\bm{r}_j)_\ell^\prime(\bm{p}_i)_k^\prime
\\
&= -(\bm{p}_i)_k\biggl((\bm{r}_j)_\ell - \frac{\epsilon\,(\bm{p}_j)_\ell+(\bm{\xi}_j)_\ell}{\alpha\,\bar{\bm{p}}^2 + \gamma} \biggr)
+ (\bm{p}_j)_\ell\biggl((\bm{r}_i)_k - \frac{\epsilon\,(\bm{p}_i)_k+(\bm{\xi}_i)_k}{\alpha\,\bar{\bm{p}}^2 + \gamma} \biggr)
\\
&= M_{ijk\ell} + \frac{(\bm{p}_i)_k(\bm{\xi}_j)_\ell - (\bm{p}_j)_\ell(\bm{\xi}_i)_k}{\alpha\,\bar{\bm{p}}^2 + \gamma} \,.
\end{split}
\eq
Hence the angular momenta are not map invariants.
Once again we sum over $i$ to obtain centroid equations
\bq
\bar{\bm{r}}^\prime = \bar{\bm{p}} \,,\qquad
\bar{\bm{p}}^\prime = -\bar{\bm{r}} - \frac{\epsilon\,\bar{\bm{p}} +\bar{\bm{\xi}}}{\alpha\,\bar{\bm{p}}^2 + \gamma} \,.
\eq
This is again effectively a single particle mapping analogous to eq.~\eqref{eq:mcmillam_map}
and it follows from eq.~\eqref{eq:foliation} that the invariant is
(the invariance is also easy to establish directly)
\bq
J_\xi = \alpha\,\bar{\bm{r}}^2 \bar{\bm{p}}^2 
+\epsilon\,\bar{\bm{r}}\cdot\bar{\bm{p}} 
+\gamma(\bar{\bm{r}}^2+\bar{\bm{p}}^2) 
+\bar{\bm{\xi}}\cdot(\bar{\bm{r}} + \bar{\bm{p}})  \,.
\eq
Hence for this model $J_\xi$ is a map invariant but regrettably the angular momenta are not conserved.
Next, let us introduce a constant vector $\bm{\beta}$ into the map equations as follows
\bq
\bm{r}_i^\prime = \bm{p}_i \,,\qquad
\bm{p}_i^\prime = -\bm{r}_i - \frac{\bm{\beta}\cdot\bar{\bm{p}}\,\bm{p}_i +\epsilon\,\bm{p}_i}
{\alpha\,\bar{\bm{p}}^2 +\bm{\beta}\cdot\bar{\bm{p}} + \gamma} \,.
\eq
Then
\bq
\begin{split}
M_{ijk\ell}^\prime &= (\bm{r}_i)_k^\prime(\bm{p}_j)_\ell^\prime - (\bm{r}_j)_\ell^\prime(\bm{p}_i)_k^\prime
\\
&= -(\bm{p}_i)_k\biggl((\bm{r}_j)_\ell - \frac{(\bm{\beta}\cdot\bar{\bm{p}} +\epsilon)\,(\bm{p}_j)_\ell}
{\alpha\,\bar{\bm{p}}^2 +\bm{\beta}\cdot\bar{\bm{p}} + \gamma} \biggr)
+ (\bm{p}_j)_\ell\biggl((\bm{r}_i)_k - \frac{(\bm{\beta}\cdot\bar{\bm{p}} +\epsilon)\,(\bm{p}_i)_k}
{\alpha\,\bar{\bm{p}}^2 +\bm{\beta}\cdot\bar{\bm{p}} + \gamma} \biggr)
\\
&= M_{ijk\ell} \,.
\end{split}
\eq
Hence the angular momenta are map invariants.
The most general form of an additional invariant, biquadratic in the coordinates and momenta
and symmetric under the interchange $\bm{r}_i \leftrightarrow \bm{p}_i$, is 
\bq
J_\beta = \alpha\,\bar{\bm{r}}^2 \bar{\bm{p}}^2 
+\lambda(\bm{\beta}\cdot\bar{\bm{r}} + \bm{\beta}\cdot\bar{\bm{p}})\,\bar{\bm{r}}\cdot\bar{\bm{p}} 
+\mu(\bm{\beta}\cdot\bar{\bm{p}}\,\bar{\bm{r}}^2 + \bm{\beta}\cdot\bar{\bm{r}}\,\bar{\bm{p}}^2)
+\epsilon\,\bar{\bm{r}}\cdot\bar{\bm{p}} 
+\gamma(\bar{\bm{r}}^2+\bar{\bm{p}}^2) \,.
\eq
Here $\lambda$ and $\mu$ are constants.
Comparing with the McMillan map, we must have $\lambda+\mu=1$.
After tedious but straightforward algebra, we obtain
\bq
\begin{split}
J_\beta^\prime 
&= J_\beta
+ \biggl[\, 2(1-\lambda)
- \frac{\lambda\,(\bm{\beta}\cdot\bar{\bm{p}} +\epsilon)}
{\alpha\,\bar{\bm{p}}^2 +\bm{\beta}\cdot\bar{\bm{p}} + \gamma} \,\biggr]\,
(\bm{\beta}\cdot\bar{\bm{p}}\,\bar{\bm{r}}\cdot\bar{\bm{p}} -\bm{\beta}\cdot\bar{\bm{r}}\,\bar{\bm{p}}^2) \,.
\end{split}
\eq
There is no value of $\lambda$ which will make the term in the brackets vanish.
Next, for $N$-body motion in one dimension, the vectors reduce to scalars and
$\bm{\beta}\cdot\bar{\bm{p}}\,\bar{\bm{r}}\cdot\bar{\bm{p}}
-\bm{\beta}\cdot\bar{\bm{r}}\,\bar{\bm{p}}^2
= \bar{\beta}\bar{p}\bar{x}\bar{p} - \bar{\beta}\bar{x}\bar{p}^2 = 0$.
In 2 or 3 space dimensions, note that
$\bm{\beta}\cdot\bar{\bm{p}}\,\bar{\bm{r}}\cdot\bar{\bm{p}}
-\bm{\beta}\cdot\bar{\bm{r}}\,\bar{\bm{p}}^2
= (\bar{\bm{p}}\times\bm{\beta})\cdot(\bar{\bm{r}}\times\bar{\bm{p}})$.
This vanishes if $\bm{\beta} \parallel \bar{\bm{r}}\times\bar{\bm{p}}$.
Note that $\bar{\bm{r}}\times\bar{\bm{p}}$ is a map invariant
because it is a sum of angular momenta (which are all individually map invariants)
and a sum of invariants is also an invariant.
However, there is a weak point in that if
$\bm{\beta} \parallel \bar{\bm{r}}\times\bar{\bm{p}}$,
then in the original map equations
$\bm{\beta}\cdot\bar{\bm{p}}=0$,
which means $\bm{\beta}$ does not appear in the map equations in the first place.
Hence an ``invariant with a term in $\bm{\beta}$'' is vacuous in 2 or 3 space dimensions.
In four or more hyperspace dimensions, 
the concept of a vector cross-product is not well-defined
(although the above dot products are well defined).
Hence for $N>1$, it is unclear that an invariant with a term in $\bm{\beta}$ exists 
for motion in more than one space dimension.
As for the model in \cite{DanilovNagaitsev_PRSTAB2014}, we can write the map equations
\bq
\bm{r}_i^\prime = \bm{p}_i \,,\qquad
\bm{p}_i^\prime = -\bm{r}_i - \frac{\bm{\beta}\cdot\bar{\bm{p}}\,\bm{p}_i +\epsilon\,\bm{p}_i}
{\alpha\,\overline{\bm{p}^2} +\bm{\beta}\cdot\bar{\bm{p}} + \gamma} \,.
\eq
The candidate invariant is (again with $\lambda+\mu=1$)
\bq
I_\beta = \alpha\,(\overline{\bm{r}\cdot\bm{p}})^2 
+\lambda(\bm{\beta}\cdot\bar{\bm{r}} +\bm{\beta}\cdot\bar{\bm{p}})\;\overline{\bm{r}\cdot\bm{p}} 
+ \mu(\bm{\beta}\cdot\bar{\bm{p}}\,\overline{\bm{r}^2} + \bm{\beta}\cdot\bar{\bm{r}}\;\overline{\bm{p}^2}) 
+\epsilon\;\overline{\bm{r}\cdot\bm{p}}
+\gamma\,(\overline{\bm{r}^2}+\overline{\bm{p}^2}) \,.
\eq
Application of the map yields
\bq
I_\beta^\prime = I_\beta +\biggl[\, 2(1-\lambda) 
-\frac{\lambda\,(\bm{\beta}\cdot\bar{\bm{p}} +\epsilon)}
{\alpha\,\overline{\bm{p}^2} +\bm{\beta}\cdot\bar{\bm{p}} + \gamma}\,\biggr]\,
(\bm{\beta}\cdot\bar{\bm{p}}\;\overline{\bm{r}\cdot\bm{p}} - \bm{\beta}\cdot\bar{\bm{r}}\;\overline{\bm{p}^2}) \,.
\eq
In this case, even for motion in one space dimension, $I_\beta$ is not invariant if $N>1$.
We must have $\bm{\beta}=0$ except for the McMillan map $N=d=1$.

To summarize, the McMillan map \cite{McMillan_map}
is a nontrivial one-dimensional rational integrable system.
A generalization to $N$-body systems (in $d$ space dimensions) was found in \cite{DanilovNagaitsev_PRSTAB2014}.
This paper presented alternative generalizations of the McMillan map to $N$-body systems.
In all cases, the models are symplectic rational integrable systems and the phase spaces are foliated 
by a suitably chosen set of angular momenta and biquadratic curves in the dynamical variables.
Additional models with a partial set of dynamical invariants were also displayed,
i.e.~not fully integrable, which demonstrate that there are nontrivial restrictions 
as to how the McMillan map can be extended to $N$-body systems in $d$ dimensions.


\end{document}